\documentclass[mathleft,fleqn,usenatbib,%
]{an}
\usepackage{graphicx}
\usepackage[varg]{txfonts}
\usepackage{mathrsfs,amssymb,amsbsy}
\usepackage{graphicx}
\usepackage{subfigure}
\usepackage{paralist}
\usepackage{color}
\usepackage{natbib}
\def\kms{\,{\rm km~s^{-1}}}
\def\kpc{\,{\rm kpc}}

\def\dex{\,{\rm dex}}
\def\Gyr{\,{\rm Gyr}}
\def\vphi{V_\phi}

\def\afe{\rm {[\alpha/Fe]}}
\def\meta{\rm {[M/H]}}
\def\feh{\rm {[Fe/H]}}

\begin{document}

\Pagespan{1}{}
\Yearpublication{2015}%
\Yearsubmission{2015}%
\Month{0}%
\Volume{999}%
\Issue{0}%
\DOI{asna.201400000}%

\title{Chemodynamics of the Milky Way  and disc formation history:\\  insight from the RAVE and Gaia-ESO surveys}

        
\author{Georges\, Kordopatis\inst{1}\thanks{On behalf of the RAVE and Gaia-ESO collaborations.}\fnmsep\thanks{Corresponding author: {gkordopatis@aip.de}} 
\and Rosemary F.\,G.\, Wyse\inst{2}
\and James Binney\inst{3}}
\titlerunning{Disc chemodynamics with RAVE and Gaia-ESO}
\authorrunning{G\, Kordopatis}
\institute{Leibniz-Institut f\"ur  Astrophysik Potsdam (AIP), An der Sternwarte 16, 14482 Potsdam, Germany
\and
Johns Hopkins University, Dept of Physics and Astronomy, 3400 N Charles Street, Baltimore, MD 21218, USA
\and
Rudolf Peierls Centre for Theoretical Physics, Keble Road, Oxford OX1\,3NP, United Kingdom}

\received{XXXX}
\accepted{XXXX}
\publonline{XXXX}


\abstract{%
Multi-object spectrographs have opened a new window on the analyses of the chemo-dynamical properties of old Milky Way stars. These analyses allow us to trace back the internal mechanisms and the external factors that have influenced  the evolution of our Galaxy, and therefore understand fundamental aspects of galaxy evolution in general. 
Here, we present recent results from the \emph{RAdial Velocity Experiment} (RAVE) and the \emph{Gaia-ESO survey}. These surveys explore the Milky Way properties in different ways, in terms of sample size and selection, magnitude range,  and spectral resolution. We focus here on (i) the first direct detection of evidence for radial migration within the thin disc, providing insight into  the history of spiral structure of the  Milky Way, 
and (ii) the chemo-dynamical characterisation of the metal-weak thick and thin discs,   for which chemo-dynamical models  still have difficulties in reproducing. }

\maketitle

\section{Introduction}
The disentanglement of the effects of past accretion events and of secular mechanisms in the evolution of the Milky Way is a complex task  \citep[e.g.][]{Rix13,Sellwood14}. Large samples of stellar velocities and metallicities are very valuable in this endeavour, containing the signatures of past events \citep[e.g.][]{Eggen62, Freeman02}, with the caveat that the distributions functions (DF)  of the different Galactic populations often overlap and change as a function of location within the Galaxy. In principle  age is the parameter with the most power to discriminate among models of  our Galaxy's history \citep[e.g.][]{Minchev14}. However,  stellar ages  have only recently started being derived with reasonable error bars, and for only  a relatively small sample of stars, and potential biases in the derived ages may still exist \citep[][]{Miglio13}. 
The large-scale investigation  of the chemo-dynamical properties of the stars constituting the different Galactic populations is therefore still the most robust way to understand  how our Galaxy was formed and evolved. This is achieved by detecting and identifying the 
signatures predicted by models that incorporate different underlying  formation mechanisms.  

These signatures are encoded in the correlations among  the spatial, chemical, and kinematic parameters of each stellar population. 
In order to detect these signatures, however, large statistical data sets are needed. For  more than a decade, multi-object spectroscopic surveys have been gathering data with continually improving spectral resolution and/or for larger number of stars. Such surveys include the Geneva Copenhagen Survey \citep{Nordstrom04}, the RAdial Velocity Experiment \citep[RAVE,][]{Steinmetz06}, the Sloan Extension for Galactic Understanding and Exploration \citep[SEGUE,][]{Yanny09}, the APO Galactic Evolution Experiment \citep[APOGEE,][]{Majewski15}, the Gaia-ESO Survey \citep[GES,][]{Gilmore12} and  the LAMOST experiment \citep{Zhao12}. The ongoing Gaia mission \citep{Perryman01} will  provide astrometric information for more than a billion stars with, in particular, measurements of the parallaxes and proper motions, in addition to photometric information and spectra for a significant fraction of the observed targets. 

Here we review some recent results, obtained with RAVE data release 4 (DR4) \citep{Kordopatis13b} and GES DR2, concerning radial migration  and the accretion history of the Galactic discs.  These results are based on analyses of the properties of the tails of the metallicity distributions of the thin and thick disc, i.e. either the super metal-rich (Sect.~\ref{sec:SMR_stars}) or the metal-poor ends (Sect.~\ref{sec:metal_poor}). Such analyses of course require large samples. Our perspectives and conclusions are presented in Sect.~\ref{sec:conclusions}.

\begin{figure*}
\includegraphics[width=\linewidth]{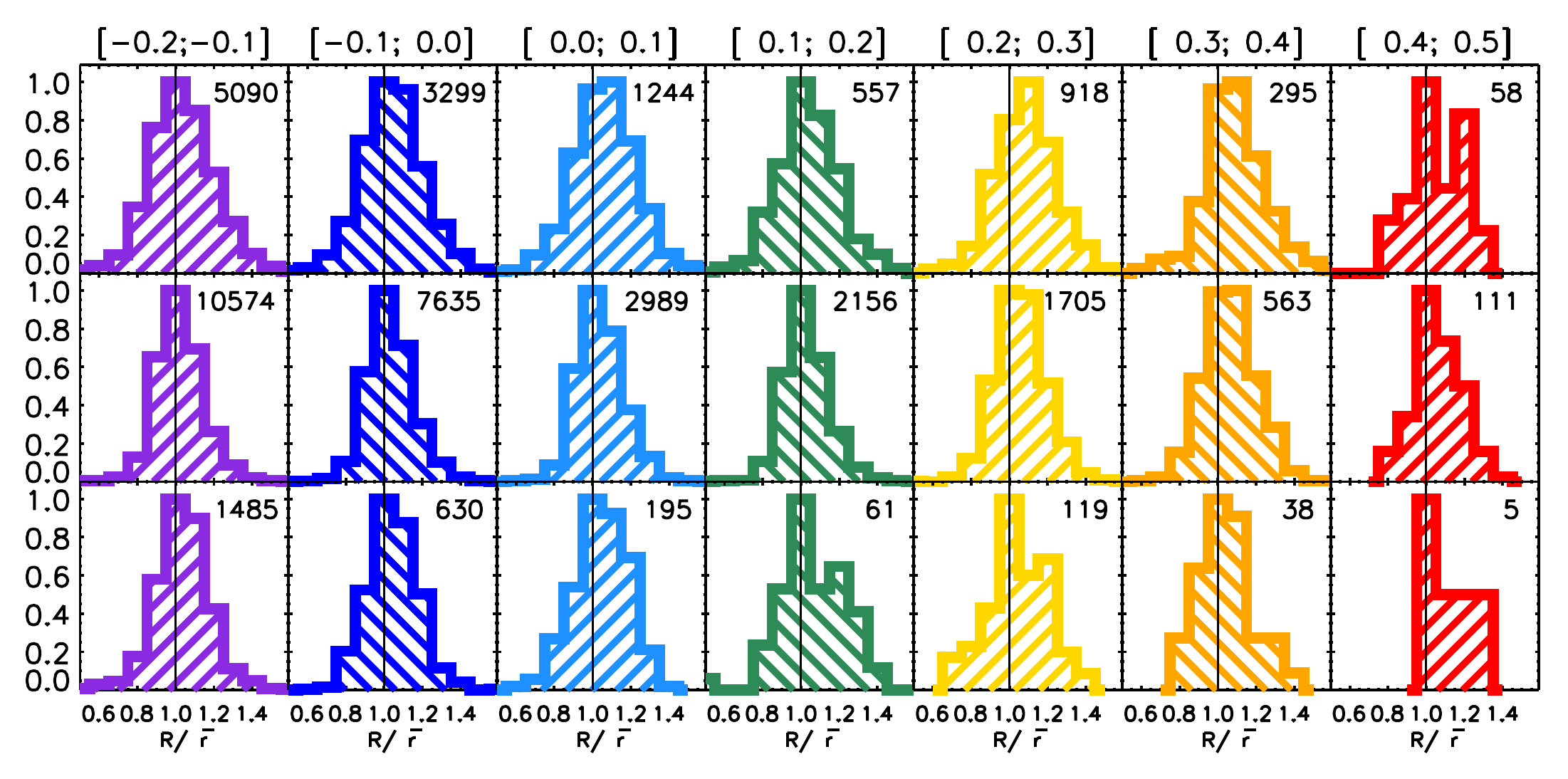}
\caption{Normalised histograms of the ratio between the observed radius and the mean orbital radius of the metal-rich RAVE stars for different metallicity bins (columns) and different Galactocentric radius rings (top:$[6.5,7.5]\kpc$, middle: $[7.5,8.5]\kpc$, bottom: $[8.5,9.5]\kpc$). The numbers in the upper right corners of each panel indicate the size of the sample used to compute that histogram.}
\label{fig:RAVE_SMR}
\end{figure*}

\section{Properties of the metal-rich end of the thin\,disc}
\label{sec:SMR_stars}
Stars in the local thin disc that have metallicities of at least twice the Solar value (denoted as  super metal-rich stars, SMR) carry important information concerning the secular evolution of the disc. Indeed, it is highly unlikely that these stars could have formed in the Solar neighbourhood, since the  local inter-stellar medium (ISM) has only now reached Solar metallicities \citep{Cartledge06}. These non-locally born stars could not have been accreted either, since a satellite galaxy containing stars of  such high metallicity is expected to have been of high mass \citep{Kirby13} and therefore its accretion should have left significant  phase-space imprints that so far have not been detected. The Super metal-rich stars have therefore been born within the disc and reached their present position either (i) through epicycle motion (therefore only passing through the Solar neighbourhood on a presumably outwards radial excursion around apocentre), or (ii) having radially migrated outwards, changing their mean orbital radius from the inner Galaxy to larger radii through interactions at the corotation resonance of transient  spiral arms and/or the central bar of the Galaxy \citep[e.g.][]{Sellwood02,Schonrich09b,Minchev10,Sellwood14}. 


The RAVE survey operated for ten years and gathered data for more than
$4\times10^5$ old FGK stars. Other than avoiding the lowest Galactic latitude fields, RAVE observed  the Southern sky homogeneously. \citet{Kordopatis15a} analysed the
spatial and orbital properties of the most metal-rich stars in RAVE's  4th
data release, and found that more than half of those with [M/H]
above 0.3 were in circular orbits, i.e. with a derived mean orbital (guiding centre) radius close to
the radial coordinate at which the star is  observed (see Fig.~\ref{fig:RAVE_SMR}). This is strong 
evidence that these stars have reached the Solar neighbourhood
through some mechanism that changed their orbital angular momentum but did not increase their orbital
eccentricities, the most likely being radial migration induced by  torques operating at the co-rotation resonance of a transient spiral
perturbation \citep[see for example,][]{Sellwood02}.

The local SMR stars are estimated to have typical ages of $\sim 6$~Gyr, similar to the Sun. Assuming a fixed radial metallicity gradient equal to that of the ISM today \citep[e.g.][]{Genovali14}, the SMR stars should have been born in the inner  regions of the Galaxy, $R\sim2\kpc$.     The results shown in Fig.~\ref{fig:RAVE_SMR} therefore highlight
that interactions between stars and transient spiral patterns at their co-rotation resonances have occurred over a significant part of  the history of the Milky Way, in order to bring a
star born  in the inner disc out to the Solar neighbourhood (its present position).  Further, such
metal-rich stars have been detected up to almost one kiloparsec above
the Galactic plane.
This puts further constraints on the history of the spiral structure in the Galaxy, since the gravitational field from the spirals should have been strong enough  to influence stars at these heights. The gravitational field of a spiral of radial wave number $k$  decreases exponentially with height above the plane, $z$, as $\exp{(-kz)}$  \citep[][]{BT08}, implying the need for spiral perturbation of small wave number (long radial wavelength) \citep[cf.][]{Solway12}. The requirement for stars to migrate over a significant fraction of the extent of the thin disc suggests these perturbations were also of high amplitude, meaning massive spiral arms  \citep[see][for further details]{Kordopatis15a}.

\section{The metal-poor ends of the discs: is there evidence of past accretion events?}
\label{sec:metal_poor}
The lowest metallicities reached by the thick and thin discs are also directly linked to the history of the Milky Way. Indeed, the low metallicity tail of the thick disc provides constraints on the  potential extragalactic origin of these stars, whereas the $\alpha$-enhancement of the low-metallicity tail of the thin disc - and how it compares with the $\alpha$-enhancement of the metal-rich tail of the thick disc -  holds information on the possible gas accretion history of the Milky Way \citep[e.g.][]{Wyse95,Chiappini97}.

\subsection{The metal-weak thick disc}
\label{sec:MWTD}
Selecting those RAVE stars located between 1 and $2\kpc$ from the
plane and with metallicities below $-1.5\dex$, \citet{Kordopatis13c}
found a homogeneously spatially distributed rotating stellar population. The
azimuthal velocity of this population was found to be correlated with
metallicity, with the amplitude of the dependence being measured as
$\partial \vphi / \partial \meta \approx 50\kms\dex^{-1}$. This equals
the correlation measured previously for stars in other surveys probing
the canonical (more metal-rich) thick disc
\citep{Kordopatis11b,Lee11,Kordopatis13a}, suggesting that this
low-metallicity population represents the metal-weak thick disc
(MWTD). The large statistical sample of RAVE allowed 
the detection of the MWTD down to metallicities as low as
$-2$~dex. Finally, despite not having accurate enough
$\alpha$-abundances to draw conclusions about their origin, the MWTD
stars in \citet{Kordopatis13c} did not show any particular distinction
compared to the halo stars of the sample. The transition between the halo and the thick disc has recently
been explicitly investigated by \citet{Hawkins15b} using APOGEE data, finding  no chemical differences between the two
populations. This confirms  earlier results based on a smaller sample
selected from RAVE, with follow-up high resolution spectroscopy
\citep{Ruchti11}.

These results highlight the smooth transition from the halo to the thick disc: the inner halo, thought to have been formed in situ \citep[e.g.][]{Carollo07, Cooper15} shows no chemical differences compared to the low-metallicity thick disc, possibly suggesting therefore that the latter was also formed in situ.  
This would be in agreement with the dynamical study of  \citet{Ruchti15}, who investigated the distribution of orbital angular momenta of the stars in the Gaia-ESO catalogue and concluded  that there was no evidence for accreted disc stars and therefore that the Milky Way  experienced a relatively quiescent merger history during the last $9-10\Gyr$ \citep[cf.][]{Wyse01}.

\begin{figure*}
\includegraphics[width=\linewidth]{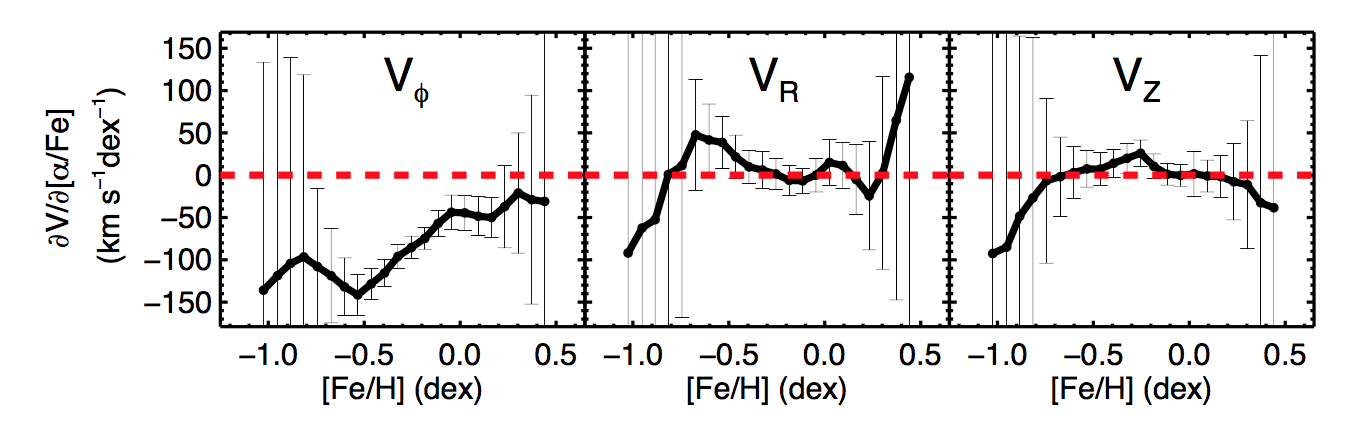}
\caption{Correlations between the $\afe$ abundances and the three velocity components for the Solar vicinity stars  ($7.5<R<8.5\kpc$ and $|Z|<1\kpc$)  of the  Gaia-ESO internal DR2. No trends are detected, within the errors, for $V_R$ and $V_Z$. In contrast, a clear signal is measured for the azimuthal velocity, indicating that the thin disc extends  down to at least $\feh\approx -0.8\dex$ and the thick disc extends up to at least Solar metallicities. }
\label{fig:GES_distinction}
\end{figure*}

\subsection{The metal-weak thin disc}

Using the second internal data-release of the high-resolution survey
Gaia-ESO\footnote{The Gaia-ESO internal DR2 comprises roughly $10^4$
stars}, \citet{Kordopatis15b} investigated, with a new approach, the 
metallicity ranges where an overlap of two populations (namely the
thin and the thick disc) could be detected. This was achieved by
determining the amplitude of  trends in the azimuthal velocities as a function of the magnesium to iron ratio (as a tracer of $\alpha$-abundances), for narrow
metallicity bins and for stars located in the extended solar
neighbourhood.  The advantage of such an analysis is that it is very
sensitive to the true tails of the metallicity distribution functions
of the underlying populations without, however, assuming any
functional shape for the MDF. Indeed, the value of the slope of the correlaton between azimuthal velocity and $\alpha$-enhancement,  $\partial \vphi / \partial \, \afe$, is expected to be non-zero  for metallicity
bins where more than a single population exist, provided that a given
stellar (mono-abundance) population  has a 
lag in mean azimuthal velocity (compared to the local standard of rest) which is
sufficiently different from that of the the other population(s) in the same sample.

The results obtained for stars located in the extended solar neighbourhood ($7.5<R<8.5\kpc$ and $|Z|<1\kpc$) are shown in Fig.~\ref{fig:GES_distinction}, where the correlations for the other velocity components are also plotted. From that figure, one can see that the trends for $V_R$ and $V_Z$ are always, within the errors, centred around zero.  This indicates that the various stellar populations present in this (sub-)sample have the same mean $V_R$ and $V_Z$ (no expansion or contraction of the populations). In contrast, the $\vphi$ panel shows clear trends, indicative of a mix of populations. 
In particular, it is found that the value of $\partial \vphi / \partial \afe$ is non-zero in the metallicity range $[-0.8, +0.2]\dex$. Furthermore, it is monotonically increasing for metallicities between  $\feh=-0.6$ and $\feh\approx 0$. These facts have been interpreted as follows: 
\begin{itemize}
\item 
The relative proportions of thin to thick disc stars changes as a function of metallicity, reflecting the difference in the mean metallicities of each population. However, there appears to be a constant fraction of thick to thin disc stars below $\feh \lesssim -0.5$, where the trend changes slope, and is consistently different than zero down to $-0.8\dex$. 
\item
Both the thin and the thick disc have highly non-Gaussian shapes for  their metallicity distribution functions. This result is in agreement with the existence of the super metal-rich tail of the thin disc (Sect.~\ref{sec:SMR_stars}) and is compatible with the existence of the metal-weak thick disc (Sect.~\ref{sec:MWTD}).
\item
The thin disc reaches metallicities as low as $-0.8\dex$. 
\item
The thick disc reaches super-solar metallicities. 
\end{itemize} 

As already noted at the beginning of Sect.~\ref{sec:metal_poor}, the $\alpha$-enhancements of the metal-weak thin disc and of the metal-rich thick disc give valuable insight into the history of gas accretion into our Galaxy. For example, the analyses of  \citet{Gilmore86}, \citet{Chiappini97} or more recently \citet{Haywood13}, \citet{Snaith14} and \citet{Bensby14},  suggest that the thick disc formed rapidly during an early, possibly turbulent, phase of the young Milky Way, and that at least part of the metal-poor thin disc formed subsequently, out of either newly 
accreted gas at the end of the star-forming epoch of the thick disc and/or from gas expelled from the thick disc. 

 
 According to these scenarios, the results derived from the Gaia-ESO data suggest two possibilities. The first is an accretion of a significant mass of metal-poor gas, to lower the ISM's metallicity before the start of the thin-disc star formation, in order to reproduce the large metallicity overlap of the discs (over $1\dex$). Alternatively, observations could also be explained with discs formed in an inside-out mode (likely combined with an violent event puffing up the thick disc), where in practice the metal-rich thick disc would have been formed in a different place than the metal-poor thin disc, without any need of decreasing the metallicity of the ISM at any radius.


\section{Conclusions and Perspectives}
\label{sec:conclusions}
The data already available allow us to gain insight into  the formation processes of the Milky Way discs and the accretion history of our Galaxy.  However, there are limitations, such as the relatively large uncertainties in distances (obtained through isochrone fitting) and proper motions (obtained from ground based surveys) leading to uncertainties in 3D  space velocities of  40-80$\kms$. In addition, individual stellar  ages are very poorly known (the relative errors from isochrone fitting is 40 percent, if not more), and the stars for which we do have a reliable  age estimate are within the Hipparcos volume and thus are very nearby and similar in metallicity to the Sun.

In the coming years, the sample for which we will have astero-seismic
ages and parallaxes and proper motion measurements from space
observations will increase significantly. This will  open new
doors into our understanding of the formation of the Milky Way, such as identifying the 
relics of minor mergers and allowing ages to be determined for  stars of
different chemical compositions. This will in turn allow us to have a ``live" picture of our Galaxy and understand the sequence of events
that have taken place during the last $10\Gyr$.

\acknowledgements
G.K.  thanks the \emph{Wilhelm und Else Heraeus-Stiftung} foundation for financial support in order to attend the ``Reconstructing the Milky Way's History: Spectroscopic Surveys, Asteroseismology and Chemodynamical Models'' meeting held in Bad Honnef in June 2015. We acknowledge the contributions of our colleagues in the RAVE and Gaia-ESO surveys.


%
%
\bibliographystyle{an}
\def\aj{AJ}\def\apj{ApJ}\def\apjl{ApJL}\def\araa{ARA\&A}\def\apss{Ap\&SS}
\def\mnras{MNRAS}\def\aap{A\&A}\def\nat{Nature}
\def\nar{New Astron. Rev.}
\bibliography{AN_Kordopatis_RW}

\end{document}